\documentclass[12pt]{article}

\usepackage{amsmath, amssymb, amsthm}
\usepackage{mathrsfs}
\usepackage{bm}

\usepackage[utf8]{inputenc}
\usepackage[T1]{fontenc}
\usepackage{lmodern}
\usepackage{microtype}
\usepackage{setspace}
\usepackage{hyperref}
\usepackage{epigraph}
\usepackage{algorithm}
\usepackage{algpseudocode}

\newtheorem{theorem}{Theorem}[section]

\newtheorem{lemma}[theorem]{Lemma}

\newcommand{\E}{\mathbb{E}}

\usepackage[
  backend=biber,
  style=phys,
  sorting=none,
  biblabel=brackets
]{biblatex}
\title{\vspace{-2cm}Lagrangians, Renormalization, \\ and Quantization in Prefix Coding}

\author{Alexander Kolpakov \footnote{Corresponding author: Alexander Kolpakov} \\ University of Austin, USA \\ Wolfram Institute, USA \\ \href{akolpakov@uaustin.org}{akolpakov@uaustin.org} 
   \and Aidan Rocke \footnote{The authors contributed equally to this manuscript} \\ Solomonoff Consulting \\ Amsterdam, The Netherlands \\ \href{rockeaidan@gmail.com}{rockeaidan@gmail.com} }

\date{\today}

\usepackage[a4paper, margin=2.5cm]{geometry}

\usepackage{charter}         

\begin{document}

\maketitle

\begin{abstract}
We develop a statistical mechanics framework for prefix coding based on variational principles, renormalization, and quantization.  
A Lagrangian formulation of entropy-optimal encoding under the Kraft--McMillan constraint yields a Gibbs-type implied distribution and completeness of the optimal code. A renormalization operator acting on codeword distribution laws produces a coarse-graining flow whose fixed points have iterated-log structure; discrete quantizations of these fixed points include Elias’ $\omega$ code as a special case.

Extending the theory to mixed discrete–continuous source laws, we show how continuous codelength functions can be quantized into countable prefix codes and derive resolution-adjusted entropy bounds together with Heisenberg-type and Boltzmann-type relations.  
This provides a unified and physically motivated view of universal coding, with Elias' $\omega$ code as a guiding example.
\end{abstract}

\section{Introduction}

Universal integer codes are a classical topic at the interface of information theory and the theory of algorithms. Elias' family of $\gamma$--, $\delta$--, and $\omega$--codes gives practical schemes whose expected lengths track entropy without prior knowledge of the source law -- a property known as \emph{universality} \cite{rissanen96}. 

The structure of Elias' $\omega$ exhibits a recursive nature: the length of an integer depends on the size of the integer and on the sizes of the integers describing that size, and so on. It can also be interpreted as scale-invariance, and thus enables a renormalization approach to the probability distribution of codelengths. 

Let $s = \text{bin}(x)$ be the binary string representation of a positive decimal number $x$, and let $x = \text{dec}(s)$ be the decimal value corresponding to a binary string $s$, possibly with leading zeros. Then Elias' $\omega$ is defined by the following two algorithms \cite{Elias1975}. 

\begin{algorithm}[H]
\caption{Elias' $\omega$: Encoding }
\label{alg:omega-encoding}
\begin{algorithmic}[1]
\Require A positive integer $N \geq 1$
\Ensure Encoded binary string for $N$

\State Initialize an empty code string $C = \text{`'}$
\State Append a "0" to the end of $C$: $C \gets C + \text{`0'}$ \Comment{Termination marker}
\While{$N > 1$}
    \State Prepend the binary representation of $N$ to $C$: $C \gets \text{bin}(N) + C$
    \State Set $N \gets \text{length}(\text{bin}(N)) - 1$ \Comment{Number of bits in $\text{bin}(N)$, minus one}
\EndWhile
\State \Return $C$
\end{algorithmic}
\end{algorithm}

\begin{algorithm}[H]
\caption{Elias' $\omega$: Decoding}
\label{alg:omega-decoding}
\begin{algorithmic}[1]
\Require Encoded binary string $S$
\Ensure Decoded positive integer $N$

\State Initialize $N \gets 1$
\State Initialize a pointer $i \gets 1$ (to track the current position in $S$)
\While{$S[i] \neq \text{`0'}$}
\State Read $N' \gets \text{dec}(S[i : i + N])$
\State Move the pointer $i \gets i + N + 1$
\State Rewrite $N \gets N'$
\EndWhile
\State \Return $N$
\end{algorithmic}
\end{algorithm}

The code description given above implies the following curious identity about its codeword lengths. 

\begin{lemma}[Elias $\omega$ codelength]\label{def:omega}
Let $n_0=n$ and recursively define $n_{j+1}=\lfloor \log_2 n_j\rfloor+1$.  
Stop at the first $m$ such that $n_m=1$.  
Then the Elias $\omega$ codelength is
\begin{equation}\label{eq:omega-def}
      \ell_\omega(n) \;=\; 1 + \sum_{j=0}^{m-1} \big(\lfloor \log_2 n_j\rfloor + 1\big).
\end{equation}
Consequently,
\begin{equation}\label{eq:omega-log}
  \ell_\omega(n)
  \;=\;
  \sum_{j=0}^{m-1} \log_2 n_j \;+\; O(\log^* n),
\end{equation}
where $\log^* n$ is the iterated base-$2$ logarithm.
\end{lemma}

The $\omega$--code is known to be \emph{prefix-free}, meaning that no codeword can represent an initial segment of another codeword. Such codes are \emph{uniquely decodable} since there is no ambiguity between the source elements encoded in a sequence and the sequence of codewords representing them.  

Being prefix-free turns out to be equivalent to the classical Kraft--McMillan \cite{Kraft1949, McMillan1956} conditions on the code word lengths. This condition turns out to be equivalent to a specific bound on the statistical sum with respect to a Gibbs measure. 

We recast Elias' $\omega$--code in a language closer to statistical physics by combining 
\begin{enumerate}
    \item[(i)] a Lagrangian formulation of optimal prefix coding under the Kraft--McMillan conditions \cite{Kraft1949,McMillan1956,cover2006elements}, and
    \item[(ii)] a renormalization map implementing coarse-graining of the codeword distribution, and normalization.
\end{enumerate}
The fixed point of the induced flow on codelengths is exactly the $\omega$--codelength; moreover, this fixed point has a wide basin of attraction.

The Lagrangian approach has been used before by Kafri to analyze Benford's law \cite{Kafri2009EntropyPI}, and by Visser \cite{Visser2013} to derive Zipf's law from the Maximum Entropy Principle \cite{jaynes1957a, jaynes1957b}. 

The renormalization approach, to the best of our knowledge, has not appeared in the earlier literature. However, it is very naturally connected to the original definition of Elias' $\omega$--code and its recursive nature. 

The encoding problem becomes a transparent toy model where scale-invariance is enforced by information-theoretical constraints (being uniquely decodable) rather than dynamics, linking two methodological pillars of statistical physics: variational principles and renormalization flows \cite{Fisher1998,Wilson1971,kardar2007fields}.

Moreover, our analysis applies to codes with both \emph{discrete} and \emph{continuous} components of their codelength distribution. The usual information constraint for a discrete code that we use is the classical Kraft--McMillan inequality that guarantees the code is uniquely decodable. Reformulating it in terms of a Gibbs measure allows writing down the continuous analogue. 

Moreover, a continuous code can be quantized by binning together some of the codewords in order to achieve a series of continuous approximations with varying \emph{resolution}. At each resolution level, the classical discrete theory applies, while This procedure recovers information-theoretical analogues of both classical Heisenberg's (resolution--complexity tradeoff in code quantization) and Boltzmann's (persistence of entropy at each resolution level) relations. 

In this work, we achieve several goals:
\begin{itemize}
    \item give a variational derivation showing that the optimizer saturates the Kraft--McMillan inequality and identifies the implied codeword probability law  -- a property otherwise called completeness (Theorem~\ref{thm:complete});

    \item quantify asymptotic optimality $\E[\ell]\le H(p)+O(1)$, with Shannon's lower bound $\E[\ell]\ge H(p)$, and interpret the discrepancies using physics analogies: code quantization, Heisenberg's and Boltzmann's relations (Theorem~\ref{thm:optimal});

    \item define a simple renormalization operator on codeword probability laws and the induced map on lengths; we prove that its fixed point is the iterated-log form of $\omega$ and that the flow is attractive up to a reasonably small error (Theorem~\ref{thm:renorm}).
\end{itemize}

\section{Optimal codes from Lagrangians and Renormalization}

\subsection{Lagrangian Optimization for Codelength}

Some interesting and practically common statistical laws appear to follow as optimal solutions of Lagrangians combining together the entropy-optimal encoding condition with various combinatorial or physical constraints. One notable example is Visser's derivation of Zipf's law in \cite{Visser2013}. Another one is Kafri's work of Benford's law \cite{Kafri2009EntropyPI} (see also \cite{KR2025}). 

Let us consider codes as probability spaces of codewords $\mathcal{C} \subseteq [1, +\infty)$, with the source distribution probability $p(x)$ of a codeword $x$ being a generalized function
\begin{equation*}
    p(dx) \;=\; \sum_{w\in W} P(w)\,\delta_w(dx) \;+\; \rho(x) dx,
\end{equation*}
i.e. a weighted sum of \emph{delta-functions} together with a continuous \emph{probability density}. Here $W$ is at most countable, $\rho\ge 0$ is integrable, $W \cap \mathrm{supp}\, \rho = \emptyset$, and
\begin{equation*}
    \sum_{w\in W}P(w)+\int_{\mathcal{C}} \rho(x)\,dx=1.
\end{equation*}

Fix an output alphabet $\Sigma_D$ of size $D\ge 2$, and let $\Sigma_D^*$ be its Kleene closure. The $D$-ary codelength $\ell_D(x)$ can be replaced by the baseless (or nat) codelength $\ell(x) = \ell_D(x) / \log D$, which we will prefer in the sequel. 

Then the average codelength equals
\begin{equation*}
    \mathbb{E}[\ell] = \int_\mathcal{C} \ell(x) p(dx). 
\end{equation*}

A \emph{prefix code} (or a \emph{prefix-free} code) is a countable set $U\subset\Sigma_D^{*}$ of finite strings over a $D$-ary alphabet such that no $u\in U$ is a prefix of another $v\in U$. 

Let us recast the Kraft--McMillan inequality \cite{Kraft1949, McMillan1956} for prefix codes in the continuous form:
\begin{equation*}
    \int_\mathcal{C} \exp(-\ell(x)) \nu(dx) \leq 1,
\end{equation*}
where
\begin{equation*}
    \nu(dx) = \sum_{w \in W} \delta_w(dx) + dx
\end{equation*}
is the \textit{Kraft measure}. 

Here, if $x \notin \mathcal{C}$ is not a codeword, we assume $\ell(x) = +\infty$. Let us note that for countable codes $p(x)$ is a linear combination of delta-functions, and the integrals above become the usual discrete sums. More details on prefix-free codes with mixed source distributions are provided in Section~\ref{sec:section-kraft}. 

We shall add this condition as a constraint to the Lagrangian below:
\begin{equation*}
    \mathcal{L}(\ell) = \int_\mathcal{C} \ell(x) p(dx) - \lambda \left( 1 - \int_\mathcal{C} \exp(-\ell(x)) \nu(dx) \right) \to \min_\ell,
\end{equation*}
with $\lambda > 0$. This implies that violating the Kraft--McMillan inequality would impose a penalty on the Lagrangian. 

The whole Lagrangian will then minimize the average codelength under the prefix-free constraint. This entropy–optimality interpretation is classical \cite{jaynes1957a,jaynes1957b,cover2006elements}.

Since $p\ll\nu$, define the Radon--Nikodym derivative
\begin{equation*}
r(x) = \frac{dp}{d\nu}(x)=
\begin{cases}
P(w), & x=w\in W,\\[4pt]
\rho(x), & x\in \mathrm{supp}\,\rho.
\end{cases}
\end{equation*}

Taking the functional derivative gives
\begin{equation*}
    \frac{\delta \mathcal{L}}{\delta \ell} = r(x) - \lambda \exp(-\ell(x)) = 0 \; (\nu\text{-a.e.}),
\end{equation*}
while 
\begin{equation*}
    1 = p(\mathcal{C}) = \int_\mathcal{C} r(x) \nu(dx) = \lambda \int_\mathcal{C} \exp(-\ell(x)) \nu(dx) = \lambda.
\end{equation*}
Indeed, the complementary slackness condition implies that the code so derived has to saturate Kraft's inequality, and thus be complete (see \cite{rockafellar70} for fundamentals of convex analysis).  

This implies $p(x)$ being the code's implied probability with
\begin{equation*}
    \ell(x) = 
\begin{cases}
- \log P(w), & x=w\in W,\\
- \log \rho(x), & x\in\mathrm{supp}\,\rho.
\end{cases}
\end{equation*}

As a result, we obtain the following theorem. 

\begin{theorem}\label{thm:complete}
    Let $\mathcal{C}$ be a self-delimiting code with source distribution probability $p(x)$ and codeword length $\ell(x)$. If $\ell(x)$ is chosen to minimize the average codelength, then the code is complete and the codeword probability is exactly the implied one. 
\end{theorem}

\subsection{Renormalizing the Codeword Distribution}

Let us define the following renormalization operator $\mathcal{R}$ for probabilities that is consistent with exponentiation. Namely, we want to be able to renormalize for $\exp(x)$ if we know the probabilities for $x$. This, on the surface, does not imply the universality property shown in \cite{Elias1975}, but rather guarantees that the conditional distribution is the same over all orders of magnitude. However, we shall see that universality will readily follow from this simple distributional condition. 

The process used below is in essence the one of coarse‐graining by replacing $x$ with $\log x$ followed by renormalizing the resulting probability (or codelength). However, it appears to be technically much simpler than the classical renormalization techniques in physics \cite{Fisher1998, Wilson1971, kardar2007fields} due to its code-theoretical nature. 

Let us put
\begin{equation*}
    \mathcal{R} p(\exp(x)) = p(x) \cdot \exp(-x), 
\end{equation*}
with the obvious fixed point given by a distribution $p(x)$ satisfying
\begin{equation*}
    p(\exp(x)) = p(x) \cdot \exp(-x).
\end{equation*}

Note that the above can also be rewritten as a statement about the codelength instead:
\begin{equation}\label{eq:renorm-1}
    \ell(\exp(x)) = \ell(x) + x.
\end{equation}
The above, together with the natural constraint $\ell(x) \geq 0$, implies that $\ell(x)$ is given by the sum of nested logarithms $\log(\log(\ldots\log(x)\ldots)$ that continues until we obtain a negative number.

This implies that 
\begin{equation}\label{eq:renorm-2}
    \ell(x) = \log x + o(\log x), \text{ as } x\to\infty.
\end{equation}

Recall that the codeword probability $p(x)$ is implied, i.e.
$r(x)=\frac{dp}{d\nu}(x)=e^{-\ell(x)}$ $(\nu$-a.e.). From equations \eqref{eq:renorm-1} -- \eqref{eq:renorm-2} we have that $\ell(x)\ge \log x$, as the additional polylog terms in the expression for $\ell(x)$ only arise while they are non-negative. Hence
\[
r(x)=e^{-\ell(x)}\le e^{-\log x}=\frac{1}{x},
\]
so $x\le 1/r(x)$ and therefore $\log x\le -\log r(x)$ pointwise. 

In turn, this yields 
\begin{equation*}
    \int_\mathcal{C} \log x\, p(dx) \leq H(p),
\end{equation*}
where 
\begin{equation*}
    H(p) = - \int_\mathcal{C} \log r(x) \, p(dx)
\end{equation*}
is the Shannon entropy \cite{Shannon1948} of $\mathcal{C}$. 

This implies that for any $\varepsilon > 0$ and all $x \in \mathcal{C}$ large enough $o(\log x) \leq \varepsilon \log x$ and thus  
\begin{equation*}
    \int_\mathcal{C} o(\log x) p(dx) \leq O(1) + \varepsilon \int_\mathcal{C} \log x\, p(dx) \leq O(1) + \varepsilon H(p) = O(1),
\end{equation*}
as $\varepsilon > 0$ can be chosen arbitrarily.

Finally, 
\begin{equation*}
    \mathbb{E}[\ell] = \int_\mathcal{C} \ell(x)\, p(dx) = \int_\mathcal{C} \log x\, p(dx) + O(1) \leq H(p) + O(1).
\end{equation*}

On the other hand, Shannon's classical lower bound $\mathbb{E}[\ell] \geq H(p)$ also holds (cf. Section~\ref{sec:section-shannon}). 

Thus, we obtain the following statement.

\begin{theorem}\label{thm:optimal}
    Let $\mathcal{C}$ be a code for which the implied probability $p(x)$ of a codeword $x \in \mathcal{C}$ satisfies the renormalization condition $p(\exp(x)) = p(x) \cdot \exp(-x)$. Then the codelength of $\mathcal{C}$ behaves as $\ell(x) = \log x + o(\log x)$, for $x\gg 1$, and $\mathcal{C}$ is asymptotically optimal: $H(p) \leq \mathbb{E}[\ell] \leq H(p) + O(1)$.  
\end{theorem}

We shall call any code satisfying Theorems \ref{thm:complete} - \ref{thm:optimal} an $\omega$--type code, for short. 

\subsection{Renormalization Flow Universality}

Let the renormalization operator \(\mathcal{R}\) act on an arbitrary codeword distribution $p(x)$ as  
\begin{equation*}
  \mathcal{R}[p](\exp (x))
  \;=\;
  p(x)\,\exp(-x),
\end{equation*}
where the map $\log: \exp(x) \to x$ implements a coarse‐graining step or scale reduction, while the factor \(\exp(-x)\) ensures 
\begin{align*}
    &\int_1^\infty \mathcal{R}[p](x)\,dx = \int^\infty_{0} \mathcal{R}[p](\exp(x)) = \\ &=  \int^\infty_{0} p(x) \exp(-x) \cdot \exp(x)dx  = \\
    &= \int^1_0 0\cdot dx + \int^\infty_{1} p(x) dx = 1,
\end{align*}
where we assume $p(x) \equiv 0$ outside of the usual codeword range $[1, +\infty)$. 

Assuming that $p(x)$ is implied, we have the induced operator for the codelength $\ell(x)$ is
\begin{equation*}
  \mathcal{T}[\ell](x)
  \;=\;
  \ell\bigl(\log x\bigr)\;+\;\log x.
\end{equation*}

Starting from any given codelength function $\ell_0(x)$, we can transform it by setting 
\begin{equation*}
    \ell_{k+1}(x) = \mathcal{T}[\ell_k](x) = \ell_k(\log x) + \log x,
\end{equation*}
for $k = 0, 1, 2, \ldots$, which gives us a discrete flow. 

The fixed point \(\ell^*(x)\) of this flow is almost the Elias' codelength from Lemma~\ref{def:omega}. Namely, we have 
\begin{equation*}
  \ell_*(x) = \mathcal{T}[\ell_*](x) = \ell_*(\log x) + \log x, 
\end{equation*}
and recursively unrolling this relation yields
\begin{equation*}
  \ell_*(x)
  =\log x
  \;+\;\log(\log x)
  \;+\;\log\bigl(\log(\log x)\bigr)
  \;+\;\cdots,
\end{equation*}
terminating when the next \(\log\) argument drops below $1$, which is directly analogous to the “0” termination of Elias' $\omega$ encoding.

If we rewrite Elias' $\omega$ codelength using natural logarithms instead of base-$2$ logarithms\footnote{i.e. we measure codelengths in nats rather than in bits.}, we get
\begin{equation*}
    \ell_\omega(x) = \ell_*(x) + O(\log^* x) = \ell_*(x) + o(\log(x)^\varepsilon),
\end{equation*}
where $\log^*(x)$ is the iterated logarithm of $x$ known for its very slow growth that guarantees the above equality for any $\varepsilon > 0$.  

Since $\log x \leq x/2$ for any $x \geq 1$, we get that
\begin{equation*}
    \ell_*(x) \leq \log x \cdot \sum^m_{i=0} \frac{1}{2^i} \leq 2 \log x 
\end{equation*}
with $m = \log^* x$. This, together with the obvious $\ell_*(x) \geq \log x$,  implies that
\begin{equation*}
    \ell_*(x) = \Theta(\log x),
\end{equation*}
all the while being a self-delimiting code unlike the usual binary encoding, which obviously also has $\Theta(\log x)$ codelength.

Moreover, given a starting non-negative function \(\ell_0(x)\) on the codewords of $\mathcal{C}$, repeated application of the map 
\begin{equation*}
    \mathcal{T} : \ell\mapsto\ell \circ \log + \log
\end{equation*}
flows to the point
\begin{equation*}
    \ell_m(x) = \sum_{i=0}^{m-1} x_i + \ell_0(x_m),
\end{equation*}
where
\begin{align*}
    x_{i+1} &= \log x_i, \text{ for } i = 0, \dots, m-1, \\
    &\text{with } x_0 = x \text{ and } x_m \leq 1. 
\end{align*}

Thus
\begin{equation*}
    |\ell_*(x) - \ell_m(x)| \leq \ell_0(x_m) = \text{const},
\end{equation*}
since $\ell_0(x_m)$ is simply the codelength of the termination marker.

\bigskip
We can summarize our findings as the following theorem. 

\begin{theorem}\label{thm:renorm}
    Let $\ell_0(x)$ be some non-negative function on the codewords of an $\omega$--type code $\mathcal{C}$. Then the repeated iterations of the renormalization map $\ell\mapsto\ell \circ \log + \log$ \emph{almost} converge to the canonical codelength. Namely, let $m = \log^* x$ and $\ell_m = \mathcal{T}^{(m)}[\ell]$. Then \[|\ell_w(x) - \ell_m(x)| = \mathrm{const} + o(\ell_\omega(x)^\varepsilon),\] for any $\varepsilon > 0$. 
\end{theorem}

\subsection{Passing to the Discrete Encoding}

Naturally, if we now recast our observation into the discrete setting, Elias' $\omega$ encoding \cite{Elias1975} will provide a practical realization with very close properties. 

As it was mentioned before,  Elias' $\omega$ encoding has to be complete. Indeed, let $\ell_\omega(n)$ be the length of $n$ in Elias' $\omega$ code, and let $\beta(n) = \lfloor \log_2 n \rfloor + 1$ be the binary length of $n$. Then we have 
\begin{equation*}
    \ell_\omega(n) = \ell_\omega(\beta(n)-1) + \beta(n).
\end{equation*}

Let also
\begin{equation*}
    I_k = \{ n \in \mathbb{N} \,|\, \beta(n) = k  \} = \{ n \in \mathbb{N} \,|\, 2^{k-1} \leq n \leq 2^k - 1\}. 
\end{equation*}

We have that $I_k$ contains $2^{k-1}$ elements, and thus 
\begin{equation*}
    S_k = \sum_{n \in I_k} 2^{-\ell_\omega(n)} = \sum_{n\in I_k} 2^{-\ell_\omega(k-1) - k} = 2^{-\ell_\omega(k-1)-1}.
\end{equation*}

Then for the Kraft's sum $S = \sum_{n\in\mathbb{N}} 2^{-\ell_\omega(n)}$ we get
\begin{equation*}
    S = \sum^\infty_{k=1} S_k = S_1 + \sum^\infty_{k=2} S_k = \frac{1}{2} + \sum^\infty_{k=2} 2^{-\ell_\omega(k-1) - 1} = \frac{1}{2} + \frac{1}{2} \sum^\infty_{k=1} 2^{-\ell_\omega(k)} = \frac{1+S}{2}.
\end{equation*}

This readily implies that $S = 1$. The real rate of convergence is very slow though: in our computations we have reached (see \cite{github-elias})
\begin{equation*}
    \sum^{2^{2^{24}}}_{n=1} 2^{-\ell_\omega(n)} \approx 0.9697265625 \ldots
\end{equation*}
 
\section{Discussion}

The case of Elias' $\omega$ encoding shows that it can be fruitful to look at an intrinsically discrete object through a continuous lens.  
Elias' $\omega$ then appears as a particular discretization or, rather, \emph{quantization} of a continuous distribution obtained as a fixed point of a flow, by replacing $x$ with the binary length $\beta(n)$ and terminating the flow when the argument drops to~$1$.

In the derivation above we never used the fact that codewords are \emph{integers} until the very last step: the renormalization flow and the fixed-point equation were all formulated for a continuous codelength function on $[1,\infty)$ endowed with a probability measure $p(dx)$.

From this point of view, the continuous formulation is not a cosmetic reformulation of a known code, but a device that explains why such a code exists and why it has a very specific mix of properties.  The Lagrangian construction identifies an implied Gibbs measure and forces completeness by saturating the Kraft–-McMillan inequality; the renormalization operator selects a fixed point whose codelength grows like an iterated sum of logarithms.  The discrete Elias $\omega$ code is then recovered as a quantized version of this continuous picture.

This perspective suggests a more general discussion continued below by introducing continuous codes and their quantization, and by deriving the corresponding Kraft inequalities, entropy bounds, and resolution–complexity tradeoffs.

\section{Continuous codes and Quantization}\label{sec:section-kraft}

\subsection{Mixed Source Distributions}

Let $p(x)$ be a probability measure on the set of codewords $\mathcal{C} \subseteq [1, +\infty)$ of the form
\[
    p(dx) \;=\; \sum_{w\in W} P(w)\,\delta_w(dx) \;+\; \rho(x) dx,
\]
where $W$ is at most countable, $\rho\ge 0$ is integrable, $W \cap \mathrm{supp}\, \rho = \emptyset$, and
\[
    \sum_{w\in W}P(w)+\int_{\mathcal{C}} \rho(x)\,dx=1.
\]

Fix an output alphabet $\Sigma_D$ of size $D\ge 2$, and let $\Sigma_D^*$ be its Kleene closure. A \emph{prefix code} (or a \emph{prefix-free} code) is a countable set $U\subset\Sigma_D^{*}$ of finite strings over a $D$-ary alphabet such that no $u\in U$ is a prefix of another $v\in U$. 

Instead of the $D$-ary codelength $\ell_D(x)$ for $x\in \mathcal{C}$, we shall use the baseless (more precisely, base-invariant) codelength $\ell(x) = \ell_D(x) / \log D$. 

In the mixed (discrete and continuous) distribution case, we shall assign source outcomes to these finite strings by grouping outcomes into finitely many \emph{coding cells} represented either by singleton atoms or quantization cells of the density. Then the notion of prefix-freeness will carry over from the standard one on the finite codebook $U$.

\subsection{Shannon's Entropy Lower Bound}\label{sec:section-shannon}

The Shannon entropy of $\mathcal{C}$ is
\[
  H(p)= - \int_{\mathcal{C}} \log\left( \frac{dp}{d\nu}(x) \right) p(dx) = -\sum_{w\in W}P(w)\log P(w)\;-\;\int_{\mathcal{C}} \rho(x)\log \rho(x)\,dx,
\]
where 
\[
\nu(dx) = \sum_{w \in W} \delta_w(dx) + dx
\]
is the Kraft measure. 

We assume that $\mathcal{C}$ satisfies the Kraft--McMillan condition:
\[
  K =\int e^{-\ell(x)}\,\nu(dx)\le 1,
\]
and call 
\[
    q(x)=\frac{e^{-\ell(x)}}{K}
\]
the \emph{Kraft distribution}. 

The latter is equivalent to  
\[
    \ell(x) = -\log q(x) - \log K, 
\]
and thus we obtain
\begin{align*}
\mathbb{E}_p[\ell] &=\int_{\mathcal{C}} \ell(x)\,p(dx)
 = - \int_{\mathcal{C}} \log q(x)\,p(dx) - \log K \\
&= H(p) + D_{\mathrm{KL}}(p\|q) - \log K \;\ge\; H(p),
\end{align*}
since $D_{\mathrm{KL}}(p\|q)\ge 0$ (by Gibbs' inequality) and $-\log K\ge 0$ (by assumption). 

Thus, we have an analogue of Shannon's lower entropy bound
\[
    \mathbb{E}_p[\ell]\ \ge\ H(p),
\]
with equality if and only if $K=1$ and $e^{-\ell}=\frac{dp}{d\nu}$ ($\nu$-a.e.). 

\subsection{Continuous Code Quantization}
Let us choose
\begin{itemize}
    \item[(i)] a finite subset $W_0\subset W$, and
    \item[(ii)] a finite measurable
partition $\Pi=\{A_1,\dots,A_m\}$ of some measurable $A\subseteq\{x:\rho(x)>0\}$.
\end{itemize}
Then $W_0$ is a \emph{set of atoms} to encode explicitly, and $\Pi$ is a \emph{quantization} of the continuous support. 

Next, let us define the finite index set
\[
  S\; =\; W_0 \,\cup\, \{A_1,\dots,A_m\},
\]
and ascribe the codelengths to its elements
\[
  \ell_s \; =\;
  \begin{cases}
    \ell(w), & s=w\in W_0,\\[2pt]
    \displaystyle - \log \int_{A_j}\exp(-\ell(x))\,dx, & s=A_j\in\Pi.
  \end{cases}
\]

\subsection{Prefix-free Quantized Codes}
A \emph{$D$-ary prefix-free quantized code} for the mixed law restricted to $S$ is a pair
$(U,\{S_u\}_{u\in U})$ where:
\begin{itemize}
  \item $U\subset\Sigma_D^{*}$ is finite and prefix-free;
  \item $\{S_u\}_{u\in U}$ is a measurable partition of the set represented by $S$, i.e. atoms in $W_0$ and cells in $\Pi$, with $\pi_u =p(S_u)>0$ for all $u\in U$.
\end{itemize}

As before, we shall use base-invariant notation or, equivalently, base $D = e$ arity. 

\textbf{Theorem.}
\emph{With $S$, $\{\pi_s\}$ and $\{\ell_s\}$ as above, we have
\[
  \sum_{s\in S} \exp(-\ell_s) \;\le\; 1,
\]
if $\mathcal{C}$ satisfies the Kraft--McMillan condition.}

Indeed, by definition,
\[
    \sum_{s\in S} \exp(-\ell_s) = \sum_{W_0} \exp(-\ell(w)) + \sum_\Pi \int_{A_j} \exp(-\ell(x)) dx \leq 
\]
\[
    \leq \sum_{W_0} \exp(-\ell(w)) + \int_\Pi \exp(-\ell(x)) dx = \int_\mathcal{C} \exp(-\ell(x))dx \leq 1.
\]  

Thus, there exists a quantized \emph{prefix} code for the mixed-distribution code $\mathcal{C}$, for every admissible choice of the index set $S$ as above.  

\subsection{Heisenberg-style and Boltzmann-style relations}

Let the Kraft sum be $K=\sum_s e^{-\ell_s}$, and let us define the
\emph{Kraft distribution} as 
\[
q_s =e^{-\ell_s}/K,
\]
which is a code-length-related Gibbs measure. 
 
Then we have the exact identity
\[
  \bar\ell  = \sum_{s\in S}\pi_s\ell_s
  \;=\; H(\pi) \;+\; D_{\mathrm{KL}}(\pi\|q) \;+\; \log\frac{1}{K}.
\]

Here $$H(\pi)=-\sum_s \pi_s\log \pi_s$$ is the Shannon entropy of the quantized source, and $$D_{\mathrm{KL}}(\pi\|q) = \sum_s \pi_s \log \frac{\pi_s}{q_s}$$ is the Kullback--Leibler (KL) divergence of two distributions \cite{cover2006elements}. 

Moreover, we can decompose $H(\pi)$ as
\[
  H(\pi)
  = H(p) - \sum_{j=1}^m \pi_{A_j}\log v_j \;+\; D_{\mathrm{KL}}(\rho\|\bar\rho),
\]
where $v_j=\int_{A_j} dx$ and $\bar\rho=\sum_j (\pi_{A_j}/v_j)\mathbf 1_{A_j}$.

Combining, we finally obtain
\[
  \bar\ell
  = \big(H(p) - \sum_{j=1}^m \pi_{A_j}\log v_j\big)
  \;+\; D_{\mathrm{KL}}(\rho\|\bar\rho)
  \;+\; D_{\mathrm{KL}}(\pi\|q)
  \;+\; \log\frac{1}{K},
\]
which implies (by Gibbs' inequality $D_{KL} \geq 0$ and Kraft's $K\leq 1$) that
\[
    \bar{\ell} \geq H(p) - \sum_{j=1}^m \pi_{A_j}\log v_j
\]

The terms of the above expressions can be interpreted as follows:
\begin{itemize}
  \item[i.] The leading term $H(p)-\sum_j \pi_{A_j}\log v_j$ is the ``resolution-adjusted entropy.''
  \item[ii.] The nonnegative correction terms reflect: within-cell mismatch $D_{\mathrm{KL}}(\rho\|\bar\rho)$, Huffman tree mismatch $D_{\mathrm{KL}}(\pi\|q)$, and Kraft slack $\log(1/K)$.
  \item[iii.] Equality holds if $\rho$ is uniform on each cell, the code is complete ($K=1$), and the Kraft distribution $q$ coincides with $\pi$.
\end{itemize}


For a quantized code $Q = (U, \{ S \}_{u\in U})$ define its \textit{resolution perplexity} as
\[
  V_Q = \exp\!\Big(\sum_j \pi_{A_j}\log v_j\Big).
\]
This plays the role of a \emph{coarse-graining scale}: the effective cell size in which the source is observed. In physics, such a scale is set by Planck’s constant when passing from classical to quantum phase space.

Since $\bar\ell$ is the average codelength in nats, then $e^{\bar\ell}$ behaves like an effective  \emph{number of distinguishable states} at the chosen resolution.
That is, $\bar \ell$ gives us the effective \textit{alphabet size}, or the number of macrostates.

All of the above implies that the inequality
\[
  \bar \ell \geq H(p) - \log V_Q,
\]
or, in the multiplicative form,
\[
    V_Q \cdot \exp(\bar{\ell}) \geq V
\]
which is directly analogous to an \emph{uncertainty principle}:
\[
  (\text{resolution}) \times (\text{complexity})
  \;\ge\; \text{information volume}.
\]

We can also write
\[
    \bar \ell \geq \log \left( \frac{V}{V_Q} \right),
\]
which means that the average codelength of the quantized code cannot be less than the information needed to specify one cell among the number of resolution-cells covering the accessible region.

Then, the optimal average codelength $\bar\ell^*$ of a $D$-ary code subject to the Kraft--McMillan inequality, satisfies a relation similar to \emph{Boltzmann's identity}. Namely,
\[
    \bar\ell^* = k_D\cdot \log W, \text{ with } W = \frac{V}{V_Q} \text{ and } k_D = \frac{1}{\log D}, 
\]
where $W$ is the reciprocal of the proportion of observable (or encodable) macrostates to the volume of entire microstate space, while $k_D$ plays the role of the ``coding Boltzmann constant''.  

\section{Conclusion}

Using Elias' $\omega$ encoding as a guiding example, we have developed a general framework that links prefix coding with ideas and tools from statistical mechanics. 

On the coding side, we start from a variational formulation of optimal codelengths under the Kraft--McMillan constraint. Then we show that the optimal solution automatically yields a complete code whose implied probability law coincides with the source distribution. 

On the physics side, this variational problem naturally produces a Gibbs-style measure on codewords and a partition-function-like Kraft sum.

We then introduce a renormalization operator acting on codeword distributions, and the associated flow on codelengths. Its fixed point exhibits the characteristic iterated-log growth familiar from Elias' $\omega$ code. 

The flow has a relatively wide basin of attraction and preserves asymptotic optimality up to an additive constant, showing that the scale-invariant structure behind Elias' construction is not an accidental but rather a robust feature of a larger family.

Finally, we extend the Kraft--McMillan framework to mixed discrete–continuous source laws and make the passage back to practical discrete codes explicit via quantization. In this setting, the average codelength decomposes into a resolution-adjusted entropy term plus non-negative corrections capturing within-cell structure, code-tree mismatch, and Kraft's slack. 

The above decomposition leads to an uncertainty-type relation that trades off resolution against coding complexity, and to a Boltzmann-style identity in which the optimal average codelength measures the logarithm of an effective number of encodable macrostates.

\section*{Acknowledgments} 

This material is based upon work supported by the Google~Cloud Research Award number GCP19980904. A.K. was also supported by the Wolfram Institute for Computational Foundations of Science, and by the John Templeton Foundation. 

\printbibliography

\end{document}